\documentclass[conference]{IEEEtran}
\usepackage[T1]{fontenc}

\usepackage{cite}

\ifCLASSINFOpdf
   \usepackage[pdftex]{graphicx}
\else
   \usepackage[dvips]{graphicx}
   \DeclareGraphicsExtensions{.eps}
\fi
\usepackage{amsmath}
\usepackage{multicol,lipsum}
\usepackage[dvipsnames]{xcolor}

\DeclareMathOperator{\sinc}{Sinc}
\DeclareMathOperator{\rect}{Rect}

\newtheorem{lemma}{\underline{Lemma}}

\usepackage[cmintegrals]{newtxmath}
\begin{document}
%
\title{Multicarrier Modulation on Delay-Doppler Plane: Achieving Orthogonality with Fine Resolutions}

\author{\IEEEauthorblockN{Hai Lin}
	\IEEEauthorblockA{Osaka Metropolitan University\\
		Sakai, Osaka, 599-8531, Japan\\
		Email: hai.lin@ieee.org}\\\vspace*{-0.8cm}
	\and
	\IEEEauthorblockN{Jinhong Yuan}
	\IEEEauthorblockA{The University of New South Wales\\
		Sydney, NSW, 2052, Australia\\
		Email: j.yuan@unsw.edu.au}\\\vspace*{-0.8cm}}

%


\maketitle

\begin{abstract}
	In this paper, we investigate the design of a novel multicarrier (MC) modulation on delay-Doppler (DD) plane, to couple the modulated signal with a doubly-selective channel having DD resolutions. A key challenge for the design of DD plane MC modulation is to find a \emph{realizable pulse} orthogonal with respect to the DD plane's fine resolutions.
	To this end, we first indicate that a feasible DD plane MC modulation is essentially a type of staggered multitone modulation. Then,
	we propose an orthogonal delay-Doppler division multiplexing (ODDM) modulation, and design the corresponding transmit pulse.
	Most importantly, we prove that the proposed transmit pulse is orthogonal with respect to the DD plane's resolutions and therefore a realizable DD plane orthogonal pulse does exist.
	Finally, we demonstrate the superior performance of the proposed ODDM modulation in terms of out-of-band radiation and bit error rate.
\end{abstract}

%

\section{Introduction}
%
%
%
%
High mobility is one of the most important scenarios of the next generation Beyond 5G/6G cellular systems, where the severely fast time-varying a.k.a doubly-selective channels make  reliable communication a very challenging task.
It is known that orthogonal frequency division multiplexing (OFDM) modulation, being adopted in the current 4G and 5G cellular systems, cannot work well in such a high mobility environment\cite{ofdm}.
In contrast to the OFDM's time-frequency (TF) plane modulation, the recently proposed orthogonal time frequency space (OTFS) modulation \cite{Hadani17,hadani2018orthogonal} 
suggests to modulate information-bearing signals on delay-Doppler (DD) plane.
The basic idea of the OTFS is to match the resolutions of the TF plane used for signal modulation to those of the TF plane used to represent the doubly-selective channel, which is modeled as a DD channel, aiming at coupling between the signal and the channel.
Then, by treating the channel's dispersive effects as potential diversity rather than undesired impairments, the OTFS can achieve better performance than the OFDM modulation in a high mobility environment.

Since the DD plane is divided with specified time (delay) and frequency (Doppler) resolutions, a modulation performed in accordance with the DD plane's resolutions, namely a DD plane modulation, is naturally a multicarrier (MC) modulation.
Obviously, a DD plane MC modulation requires a pulse orthogonal with respect to the DD plane's resolutions. For the sake of conciseness, in this paper, we call the pulse orthogonal with respect to the DD (or TF) plane's resolutions as DD (or TF) plane orthogonal pulse, and use TF (or DD) plane and TF (or DD) domain interchangeably. Also, without special notice or explanation, the TF plane refers to the signal plane with \emph{coarse} resolutions adopted in the OFDM modulation.

Obviously, a pulse confined to one fine grid of the DD plane is  a DD plane orthogonal pulse. However, according to the uncertainty principle, this particular DD plane orthogonal pulse doesn't exist.
In fact, the OTFS modulation may be considered as a practical workaround for this difficulty \cite{Hadani17}.
The OTFS's DD domain signal is first mapped to a TF domain signal via the inverse symplectic finite Fourier transform (ISFFT), and then conveyed by the TF plane rectangular pulse, which is essentially a TF plane orthogonal pulse \cite{Hadani17}.
In other words, the OTFS signal is still orthogonal with respect to the TF plane's coarse resolutions, and its ideal pulse is said to satisfy the \emph{TF plane bi-orthogonal robust property} with respect to the time and frequency translations induced by the doubly-selective channel. Unfortunately, such a TF plane ideal pulse cannot be realized in practice\cite{hadani2018orthogonal}. To achieve the coupling between the modulated signal and the DD channel, a DD plane MC modulation is a nature and better choice.
However, to the best of our knowledge, currently there is no DD plane MC modulation, because whether a realizable DD plane orthogonal pulse exists or not, is still unknown.

In this paper, we investigate the DD plane MC modulation, and answer the above fundamental question for the DD plane MC modulation design.
We propose a novel orthogonal delay-Doppler division multiplexing (ODDM) modulation, and design the corresponding transmit pulse. Most importantly, we prove that the proposed transmit pulse is orthogonal with respect to the DD plane's resolutions and therefore a realizable DD plane orthogonal pulse does exist. We show that the proposed ODDM is a DD plane MC modulation and can achieve \emph{perfect coupling} between the modulated signal and the DD channel. Our contribution can be summarized as follows:

\begin{figure*}
	\centering
	\includegraphics[width=0.98\textwidth]{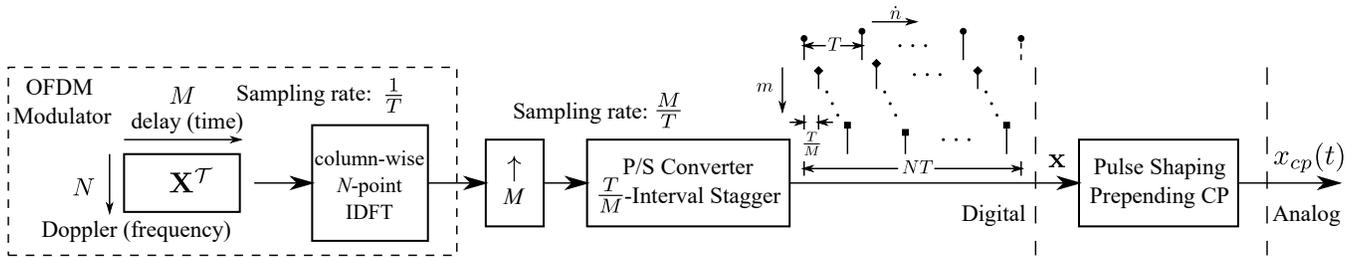}
	\caption{ODDM modulation}
	\label{oddm}
	\vspace{-4mm}
\end{figure*}

\begin{itemize}

	\item By clarifying that the time resolution on signal plane is symbol interval rather than symbol period, we indicate that a feasible DD plane MC modulation is essentially a type of the staggered multitone (SMT) modulation\cite{mct}.

	\item Analogous to the OFDM, we propose an ODDM modulation and present it as a staggered upsampled-OFDM in the digital domain. We also indicate that the ODDM can be viewed as a pulse-shaped OFDM (PS-OFDM) and propose the corresponding transmit pulse.

	\item We prove the \emph{orthogonality} of this particular transmit pulse with respect to the DD plane's resolutions, and show that the proposed transmit pulse is exactly the desired DD plane orthogonal pulse for DD plane MC modulation.

	\item The associated DD domain channel input-output relation of the proposed ODDM modulation is derived by directly exploiting the well-known frequency domain properties of OFDM symbol with timing and frequency offsets. We show that the equivalent channel matrix has an elegant block-circulant-like structure.

	\item The superior performance of the proposed ODDM modulation over the OTFS is confirmed from performance comparisons in terms of out-of-band (OOB) radiation and bit error rate (BER).

\end{itemize}

\section{ODDM modulation}

\subsection{TF and DD planes}
Suppose that a doubly-selective channel is composed of $\tilde P$ paths.
Passing a band-limited and time-limited signal through the doubly-selective channel, we observe a corresponding \emph{band-limited and time-limited} $P$-path ($P\ge \tilde P$) equivalent DD channel with delay resolution $\frac{1}{B_0}$ and Doppler resolution $\frac{1}{T_0}$, where $B_0$ and $T_0$ are the sampling rate and the signal duration, respectively.
Let $l_p, k_p \in \mathbb Z$, the equivalent DD channel is given by \cite{bello}
\begin{align}\label{ddchannel}
	h(\tau,\nu)=\sum_{p=1}^{P} h_p\delta(\tau-\tau_p)\delta(\nu-\nu_p),
\end{align}
where $\tau$ represents delay variable, $\nu$ represents Doppler variables, and $h_p$, $\tau_p=l_p\frac{1}{B_0}$, and $\nu_p=k_p\frac{1}{T_0}$ are the gain, delay, and Doppler of the $p$th path, respectively.

Let us first consider an OFDM-modulated signaling, we have the following grid consideration on the TF plane:
\begin{itemize}
	\item TF plane grid $\Pi$ :  $\{\dot nT,\dot m\frac{1}{T}\}$ for $\dot n=0,\ldots,N-1$ and $\dot m=0,\ldots M-1$, where $N$ refers to the number of OFDM symbols, and $M$ refers to the number of subcarriers for each OFDM symbol. In addition, each OFDM symbol has the duration of $T$, and $\frac{1}{T}$ is the sub-carrier spacing of the OFDM symbol.
\end{itemize}
For the OFDM signals, we have $B_0=\frac{M}{T}$ and $T_0=NT$. Therefore, the DD channel's resolutions are $\frac{T}{M}$ and $\frac{1}{NT}$, respectively.
Comparing the TF plane grid $\Pi$ for the OFDM to the DD channel's resolutions, it is clear that there is a resolution mismatch between the OFDM signal and the doubly-selective channel, which will cause complicated inter-symbol interference (ISI) and inter-carrier interference (ICI), and then subsequently severe performance degradation. To deal with this problem, we propose a DD plane modulation by modulating signals on the following DD plane
\begin{itemize}
	\item DD plane grid $\Gamma $ :  $\left\{m\frac{T}{M}, n\frac{1}{NT}\right\}$ for $m=0,\ldots,M-1$ and $n=0,\ldots N-1$, where $m$ and $n$ denote the $m$-th delay and $n$-th Doppler, respectively.
\end{itemize}
to couple the modulated signal with the DD channel having fine resolutions.

\subsection{ODDM digital sequence}
One can see that the DD plane is just a TF plane with \emph{fine} grids corresponding to the delay and Doppler resolutions, where
an MC modulation requires a corresponding DD plane orthogonal pulse. For MC modulation, a common sense\cite{mct} so far is that the area of grid is greater than or equal to $1$, for example, we have $T\times \frac{1}{T}=1$ for the TF grid $\Pi$ corresponding to a realizable transmit pulse.
On the other hand, we have $\frac{T}{M}\times \frac{1}{NT}=\frac{1}{MN} \ll 1$ for the DD grid $\Gamma$, where the obvious DD plane orthogonal pulse confined to $\Gamma$ violates the uncertainty principle
and therefore cannot be realized. As a result, it seems impossible to achieve an MC modulation on the DD plane.

Here, we would like to clarify an important concept in the context of MC modulation, that  the time resolution on signal plane is \emph{not symbol period but symbol interval}. In the meantime, the frequency resolution or the frequency spacing is the inverse of symbol period. The tiny grid of the DD plane implies that the symbol period is longer than the symbol interval, and therefore successive MC symbols are staggered. Hence, a feasible DD plane MC modulation is essentially a type of the SMT modulation\cite{mct}.

Leaving the question about the existence of realizable DD plane orthogonal pulse for the time being, we first propose an ODDM modulation, which is an orthogonal MC modulation that modulates signals in accordance with the DD plane's  delay and Doppler resolutions. Recall that $\Gamma=\left\{m\frac{T}{M}, n\frac{1}{NT}\right\}$ for $m=0,\ldots,M-1$ and $n=0,\ldots N-1$, the proposed ODDM is generated in a similar way as an $N$-subcarrier OFDM, whose subcarrier spacing, symbol period, and symbol interval are $\frac{1}{NT}$, $NT$, and $\frac{T}{M}$, respectively.
Because the symbol interval is $\frac{T}{M}$, the total bandwidth of ODDM will be around $\frac{M}{T}$, rather than $\frac{1}{T}$ in the conventional $N$-subcarrier OFDM with subcarrier spacing $\frac{1}{NT}$.
To obtain the ODDM, the $N$ discrete samples of one OFDM symbol need to be upsampled by $M$.
Fig. \ref{oddm} shows the block diagram of the ODDM. Let an $M\times N$ matrix $\mathbf X$ consisting of $MN$ transmit quadrature amplitude modulation (QAM) symbols, the ODDM in digital domain can be represented as a staggered upsampled-OFDM in Fig. \ref{oddm}, where $\mathcal T$ denotes the transpose operator, while the upsampling factor and the stagger interval are $M$ and $\frac{T}{M}$, respectively.

\begin{figure}
	\centering
	\includegraphics[width=8.8cm]{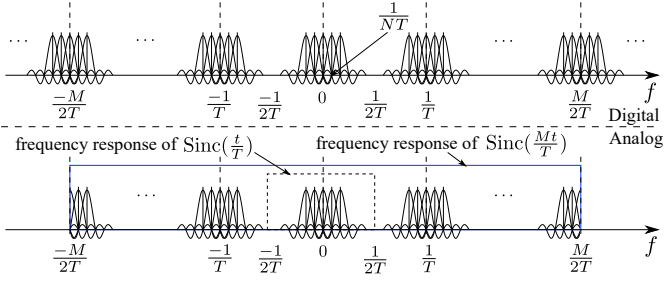}
	\caption{Signal spectrum before and after interpolation}
	\label{su-ofdm-spec}
	\vspace{-6mm}
\end{figure}

\vspace{-1mm}

\subsection{Pulse shaping and ODDM waveform}
For $N$ subcarriers spaced by $\frac{1}{NT}$ to carry $N$ QAM symbols ${\boldsymbol S}=[S_0,\ldots, S_{N-1}]^\mathcal T$, let $\mathbf s=[s_0,\ldots, s_{N-1}]^\mathcal T$ be the $N$-point inverse discrete Fourier transform (IDFT) of $\boldsymbol S$.
Since in the OFDM modulation, each symbol is one cycle of a periodic signal, the pulse shaping of OFDM can be performed by first interpolating
the cyclically extended
$\mathbf s$ using the ideal interpolation filter (IIF) $\sinc(\frac{t}{T})$ to generate $\tilde s(t)= \sum_{n=-N/2}^{N/2-1} S_{[n]_N} e^{j2\pi\frac{n}{NT}t}$, where $[\cdot]_N$ stands for the mod $N$ operator.
Then, a transmit pulse $g_{tx}(t)$ based windowing is applied to $\tilde s(t)$ to obtain the analog OFDM symbol $s(t)=\sum_{n=-N/2}^{N/2-1} S_{[n]_N} g_{tx}(t) e^{j2\pi\frac{n}{NT}t}$. Because the transmit pulse in the conventional OFDM is an $NT$-length rectangular pulse $\rect_{NT}(t)$ whose Fourier transform is $NT\sinc(fNT)$, $s(t)$ is only \emph{barely} banded to $\left(-\frac{1}{2T}, \frac{1}{2T}\right)$. Therefore, in a practical OFDM system, some edge subcarriers are unloaded to not only sharp the spectrum but also ease the
filtering at the transceiver. Meanwhile, once these edge subcarriers are unloaded and $s(t)$ can be treated as \emph{essentially} bounded to $\left(-\frac{1}{2T}, \frac{1}{2T}\right)$, the pulse shaping can be approximately simplified to interpolating $\mathbf s$ using $\sinc(\frac{t}{T})$, by treating $\mathbf s$ as the Nyquist sampling result of $s(t)$.

In the proposed ODDM modulation, we do upsample each digital MC symbol by $M$ to stagger them by the stagger interval $\frac{T}{M}$. In other words, a corresponding transmit pulse $g_{tx}(t)$ is required to guarantee ISI-free among $M$ staggered MC symbols with $N$ orthogonal subcarriers, and hence each MC symbol becomes a PS-OFDM symbol.
Since the bandwidth is increased $M$ times by the upsampling, it is natural to consider what happens to the signal spectrum if we use the \emph{wideband} IIF $\sinc(\frac{Mt}{T})$ to generate an analog symbol.
Note that although it may not be a Nyquist sampling, a $\frac{1}{T}$-rate sampling still can be used to obtain $N$ samples of an OFDM symbol from its analog version, for example, obtaining $\mathbf s$ from $s(t)$ when $g_{tx}(t)=\rect_{NT}(t)$.
Then, as shown in Fig. \ref{su-ofdm-spec}, the aliasing caused by the $\frac{1}{T}$-rate sampling spreads the spectrum of the analog OFDM symbol
over the frequency axis, where several edge subcarriers are not plotted and the shape of $NT\sinc(fNT)$ is truncated for display purpose. After passing the digital samples through the ideal interpolation filter $\sinc(\frac{Mt}{T})$,
the signal spectrum is bounded to $\left(-\frac{M}{2T}, \frac{M}{2T}\right)$, where the conventional OFDM signal spectrum roughly bounded to $\left(-\frac{1}{2T}, \frac{1}{2T}\right)$ is also pointed out for comparison.

\begin{figure}[t]
	\centering
	\includegraphics[width=8.8cm]{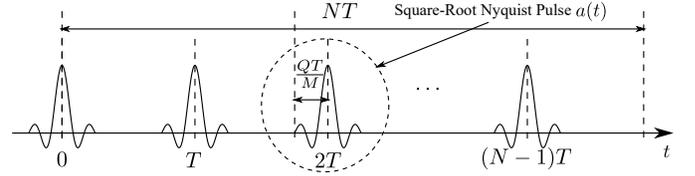}
	\caption{Transmit pulse $u(t)$}
	\label{utwave}
	\vspace{-6mm}
\end{figure}

Notice that $\sinc(\frac{Mt}{T})$ is exactly an ISI-free pulse with interval $\frac{T}{M}$, the signal spectrum in Fig. \ref{su-ofdm-spec} inspires us that the ODDM's transmit pulse should be Nyquist with interval $\frac{T}{M}$, and at the same time have a period of $NT$. In other words, the pulse is \emph{locally wideband and globally narrow-band}.
Bearing in mind that $M$ symbols are staggered, let us consider a time-symmetric real-valued square-root Nyquist pulse $a(t)$
with a time duration of $2Q\frac{T}{M}$, where $Q \ll \frac{M}{2}$ and $\int_{-\infty}^{+\infty}|a(t)|^2 dt=\frac{1}{N}$.
We then propose to use
\vspace{-1mm}
\begin{align}\label{ut}
	u(t)=\sum_{\dot n=0}^{N-1}a(t-\dot nT),
\end{align}
a pulse train shown in Fig. \ref{utwave}, as the transmit pulse $g_{tx}(t)$ for the ODDM, where $\int_{-\infty}^{+\infty}|u(t)|^2 dt=1$.

In the context of PS-OFDM, the analog OFDM waveform with the transmit pulse $u(t)$ can be written as
$\tilde x_m(t) = \sum_{n=0}^{N-1} X(m,n) e^{j2\pi\frac{n t}{NT}}  u(t)$,
where $X(m,n)$ is the signal at the $n$-th subcarrier of the $m$-th symbol and $N$ subcarriers are shifted to positive frequencies for simplicity of notation. We then have $M$ staggered $N$-subcarrier OFDM symbols pulse-shaped by $u(t)$ to form an ODDM frame
\begin{align}\label{xtut}
	x(t)  =  \sum_{m=0}^{M-1}\sum_{n=0}^{N-1} X(m,n) u\left(t-m\frac{T}{M}\right) e^{j2\pi\frac{ n }{NT}(t-m\frac{T}{M})},
\end{align}
where $-Q\frac{T}{M} \le t \le NT+Q\frac{T}{M}$. Furthermore, due to the channel delay spread,
an $(L-1)\frac{T}{M}$-length of CP is appended to the head of the frame.
Considering the added CP, we can extend the definition of $u(t)$ to
$u_{cp}(t)=\sum_{\dot n=-1}^{N-1}a(t-\dot nT)$,
then the CP-included ODDM waveform spanning over $-(L+Q-1)\frac{T}{M} \le t \le NT+Q\frac{T}{M}$ becomes
\begin{eqnarray}\label{xtcp}
	x_{cp}(t) = \sum_{m=0}^{M-1}\sum_{n=0}^{N-1} X(m,n) u_{cp}\left(t-m\frac{T}{M}\right) e^{j2\pi\frac{ n }{NT}(t-m\frac{T}{M})},
\end{eqnarray}
where $u_{cp}(t)=u(t)$ for $t\in \left(-Q\frac{T}{M}, (N-1)T+Q\frac{T}{M}\right)$ and $u_{cp}(t)=0$ for $t\in \left(-T+Q\frac{T}{M}, -Q\frac{T}{M}\right)$.
For $u(t)$, we have the following lemma,
\begin{lemma}\label{l1}
	$u(t)$ satisfies the orthogonal property that
	\begin{equation}
		A_{u,u}\left(m\frac{T}{M}, n\frac{1}{NT}\right)= \delta(m)\delta(n),
		\label{orth}
	\end{equation}
	for $|m|\le M-1$ and $|n|\le N-1$.
\end{lemma}
\begin{IEEEproof}
	See Appendix A.
\end{IEEEproof}
As a result, $u(t)$ is exactly the DD plane orthogonal pulse that we are looking for.

\section{ODDM Demodulation}

After receiving an ODDM frame distorted by the doubly-selective channel, the receiver performs the matched filtering to receive the transmitted signal.
The orthogonality of $u(t)$ makes it fulfill the perfect reconstruction condition with respect to $\Gamma$.
Also, because the DD channel's delay and Doppler are \emph{integer multiples} of delay resolution and Doppler resolution, the deviation of the channel input-output relation becomes significantly simple.

\subsection{DD domain channel input-output relation}
From (\ref{xtcp}), we have the received signal for $-(L+Q-1)\frac{T}{M} \le t \le NT+(L+Q-1)\frac{T}{M}$ as
\begin{align}
	y(t) & = \sum_{p=1}^{P} h_p x_{cp}(t-\tau_p) e^{j2\pi \nu_p (t-\tau_p)}+z(t), \nonumber                                \\
	     & = \sum_{p=1}^{P} \sum_{m=0}^{M-1}\sum_{n=0}^{N-1} h_p X(m,n) u_{cp}\left(t-(m+l_p)\frac{T}{M}\right)  \nonumber \\
	     & \quad \times e^{j2\pi \frac{(n+k_p)}{NT}(t-(m+l_p)\frac{T}{M})} e^{j2\pi\frac{k_p m}{MN}}+z(t).
\end{align}
Because of $u_{cp}(t)=u(t)$ for $t\in \left(-Q\frac{T}{M}, (N-1)T+Q\frac{T}{M}\right)$,  after the matched filtering based on $u(t)$, we obtain the signal at the $n$-th subcarrier of the $m$-th symbol as
\begin{align}
	Y(m,n) & = \int y(t) u\left(t-m\frac{T}{M}\right) e^{-j2\pi\frac{ n }{NT}(t-m\frac{T}{M})} dt, \nonumber         \\
	       & = \sum_{p=1}^{P} h_p \tilde X(\dot m,\dot n) e^{j2\pi\frac{k_p (m-l_p)}{MN}}+z(m,n),   \label{DDsignal}
\end{align}
where $\tilde X(\dot m,\dot n) = X(\dot m,\dot n)$ for $\dot n =[n-k_p]_N$ and $\dot m = m-l_p\ge 0$, and $z(m,n)$ is the DD domain noise sample.
When $\dot m < 0$, because of the CP,
the $\dot m$-th symbol is just a $T$ cyclic time-shift of the $[\dot m]_M$-th symbol.  Since cyclically time-shifting a PS-OFDM symbol with a subcarrier spacing of $\frac{1}{NT}$ by $T$ corresponds to a phase rotation term $e^{-j2\pi\frac{\dot n}{NT}T}=e^{-j2\pi\frac{\dot n}{N}}$ applied to its frequency domain signal, for $\dot m < 0$ in (\ref{DDsignal}),
we have $\tilde X(\dot m,\dot n) = e^{-j2\pi\frac{\dot n}{N}} X([\dot m]_M,\dot n)$.

The sample-wise result in (\ref{DDsignal}) can be vectorized to obtain a more insightful symbol-wise DD channel input-output relation.
Without loss of generality, assume that the maximum delay and Doppler of the channel are  $(L-1)\frac{T}{M}$ and $K\frac{1}{NT}$, respectively.
The $P$ paths can be arranged in a $(2K+1)\times L$ DD domain channel matrix $\mathbf G$, where each row and column of $\mathbf G$ correspond to a Doppler and a delay index, respectively. Clearly, the total number of non-zero elements in $\mathbf G$ is $P$.

Similar to $\mathbf x_m=[X(m,0),\ldots, X(m,N-1)]^{\mathcal T}$, namely the $m$-th column of $\mathbf X^{\mathcal T}$, we can use an $N\times 1$ vector $\mathbf y_m=[Y(m,0),\ldots, Y(m,N-1)]^{\mathcal T}$ to represent the frequency (i.e. Doppler) domain signal of the $m$-th received ODDM symbol.
Seen from the $m$-th received ODDM symbol, the path with a delay of  $l\frac{T}{M}$ brings an ISI from the $(m-l)$-th ODDM symbol, where the path's Doppler $\hat k\frac{1}{NT}$ cyclically shifts the subcarrier of the interfering $(m-l)$-th OFDM symbol by $\hat k=k-K-1$. Also, since the $(m-l)$-th ODDM symbol starts for $\frac{ (m-l)T}{M}$, the Doppler also introduces a phase rotation $e^{j2\pi\hat k\frac{1}{NT}\frac{ (m-l)T}{M}}=e^{j2\pi\frac{\hat k(m-l)}{MN}}$. As a result, for the $m$-th received ODDM symbol, the ISI from the $(m-l)$-th ODDM symbol, which is introduced by all paths with the same delay of $l\frac{T}{M}$ but different Dopplers, can be governed by
$\mathbf H_l^m  =  \sum_{k=1}^{2K+1} g(k,l) e^{j2\pi \frac{\hat k(m-l)}{MN} } \mathbf C^{\hat k}$,
where $\mathbf C$ is the $N \times N$ cyclic permutation matrix.

From the above analysis, we know that for each $\mathbf y_m$, the signal term from $\mathbf x_{m-l}$ is $\mathbf H_l^m \mathbf x_{m-l}$, for $0\le l \le L-1$. When $m-l<0$, like (\ref{DDsignal}),
additional phase rotation term $\mathbf D$ is applied to $\mathbf x_{[m-l]_M}$, where
$\mathbf D = \textrm{diag}\{1, e^{-j\frac{2\pi}{N}},\ldots, e^{-j\frac{2\pi(N-1)}{N}}\}$.
Therefore, let $\mathbf z_m$ denote the $m$-th noise vector, the input-output relation in the DD domain can be written in a matrix form as
\begin{equation} \label{ioddcompact}
	\mathbf y=\mathbf H \mathbf x + \mathbf z,
\end{equation}
where
$\mathbf y  =  [\mathbf y_0^{\mathcal T},\mathbf y_1^{\mathcal T}, \cdots, \mathbf y_{M-1}^{\mathcal T} ]^{\mathcal T}$, $\mathbf x  =  [\mathbf x_0^{\mathcal T},\mathbf x_1^{\mathcal T}, \cdots, \mathbf x_{M-1}^{\mathcal T} ]^{\mathcal T}$,
$\mathbf z  =  [\mathbf z_0^{\mathcal T},\mathbf z_1^{\mathcal T}, \cdots, \mathbf z_{M-1}^{\mathcal T} ]^{\mathcal T}$, and $\mathbf H$ is the equivalent DD domain channel with size $MN \times MN$ given by
\begin{equation}\label{H}
	\mathbf H=
	\begin{bmatrix}
		\mathbf H_0^0         &        &        &                       & \mathbf H_{L-1}^0 \mathbf D & \cdots         & \cdots & \mathbf H_{1}^0  \mathbf D      \\
		\vdots                & \ddots &        &                       &                             & \ddots         & \ddots & \vdots                          \\
		\vdots                & \ddots & \ddots &                       &                             &                & \ddots & \vdots                          \\
		\mathbf H_{L-2}^{L-2} & \ddots & \ddots & \mathbf H_{0}^{L-2}   &                             & \mbox{\Huge 0} &        & \mathbf H_{L-1}^{L-2} \mathbf D \\
		\mathbf H_{L-1}^{L-1} & \ddots & \ddots & \ddots                & \mathbf H_{0}^{L-1}         &                                                           \\
		                      & \ddots & \ddots & \ddots                & \ddots                      & \ddots                                                    \\
		                      &        & \ddots & \ddots                & \ddots                      & \ddots         & \ddots                                   \\
		\mbox{\Huge 0}        &        &        & \mathbf H_{L-1}^{M-1} & \ddots                      & \ddots         & \ddots & \mathbf H_{0}^{M-1}
	\end{bmatrix}.
\end{equation}

For sparse channels, most of $\mathbf H_l^m$ are zero matrices. While for those non-zero $\mathbf H_l^m$, if there is only one path at that delay $l$, $\mathbf H_l^m$ is simply an $N\times N$ cyclic shift permutation matrix up to a scale factor. Moreover, even there are multiple paths with different Dopplers at the same delay $l$, $\mathbf H_l^m$ is still an $N\times N$ circulant matrix.
Meanwhile, it can be observed that regardless of the channel sparsity, the channel matrix $\mathbf H$ in (\ref{H}) has an elegant block-circulant-like structure, which can be exploited in signal detection.

\subsection{Signal detection}
From (\ref{ioddcompact}) and (\ref{H}), it is clear that an effective data detector is required to unlock the full time and frequency diversity potentials offered by ODDM in order to obtain reliable error performance.
Considering that the main focus of the paper is a novel DD plane MC scheme, we employ a commonly deployed DD domain message-passing (MP) detector \cite{8424569,ampotfs},
where the total interference and noise at each observation node is assumed to follow a Gaussian distribution.
The detector's computational complexity is in the order of $MNP$, where $P$ is also the number of nonzero entries in each row of the DD domain channel matrix in (\ref{H}).

\section{Remarks on TF signal distribution}
For a signal with bandwidth $B_0$ and duration $T_0$, it is well-known that its dimension or degrees of freedom (DoF) is approximately $B_0T_0$ \cite{fwc}. As a result, to transmit $B_0T_0=MN$ complex QAM symbols, we essentially need a TF region whose area is not less than $MN$. On the other hand, the $MN$ DD plane grids in Section II only occupy an area of $MN\times \frac{1}{MN}=1$, and therefore there is no way to transmit $MN$ QAM symbols if we are limited to this small TF region.
In other words, \emph{not only the TF resolutions but also the totally occupied TF region} should be considered in the design of a modulation.

Since different TF resolutions correspond to different TF signal distributions inside the total TF region, which is bounded by the DoF, it is meaningful to compare the TF signal distribution of the ODDM to that of the OTFS. Due to the space limitation, only a brief summary of the comparison is given below.
For OTFS, the DD domain signal is mapped to the TF domain via the ISFFT, therefore its TF signal distribution is just that of the OFDM. As a result, without an ideal pulse, the OTFS suffers from the blurred ISI and ICI caused by the resolution mismatch.
On the other hand, by use of $u(t)$, the ODDM staggers $M$ symbols with an interval of $\frac{T}{M}$.
Also, it can be shown that each ODDM symbol has a spectrum similar to that in Fig. \ref{su-ofdm-spec}, where the $N$ subcarriers spaced by $\frac{1}{NT}$ are spread $M$ times, to form a signal structure equivalent to having \emph{cyclic prefix and suffix in frequency domain}. Therefore, the TF plane in the ODDM is actually ``oversampled" to a DD plane to perform modulation in a 2D uniformly distributed fashion, achieving an orthogonality with fine resolutions. Consequently, the ODDM enjoys the prefect coupling between the modulated signal and the DD channel and only experiences the well-controlled on-the-grid ISI and ICI introduced by the channel. In particular, the cyclic signal structure in both time and frequency domains results in a block-circulant-like DD domain channel matrix $\mathbf H$ in (\ref{H}).

It is also noteworthy that the ODDM and the OTFS have the same time domain digital sequence shown in Fig. \ref{oddm}. When $Q\ll \frac{M}{2}$, it can be proved that the $u(t)$-based pulse shaping for the ODDM can be approximated by an $a(t)$-based filtering\cite{oddm}. Therefore, a digital OTFS signal with an $a(t)$-based filtering approximates an ODDM waveform.

\section{Simulation Results}
In this section, simulations are conducted to verify the performance of the proposed ODDM modulation, especially compared to the OTFS modulation. The simulation parameters are: carrier frequency $5$GHz, $\frac{1}{T}= 15$kHz, CP length $3.125\mu$s.
For the channel, similar to \cite{8424569}, we adopt the Extended Vehicular A (EVA) model, where each path has a single Doppler generated using Jake's formula $\nu_p = \nu_{\textrm{max}} \cos(\theta_p)$, the maximum Doppler $\nu_{\textrm{max}}$ is determined by the user equipment (UE) speed and $\theta_p$ is uniformly distributed over $[-\pi,\pi]$.
For the ODDM, $a(t)$ is a square-root raised cosine pulse with roll-off factor $0.25$ and $Q=16$, while for the OTFS, the TF plane rectangular pulse is employed.

\begin{figure}
	\centering
	\includegraphics[width=8.8cm]{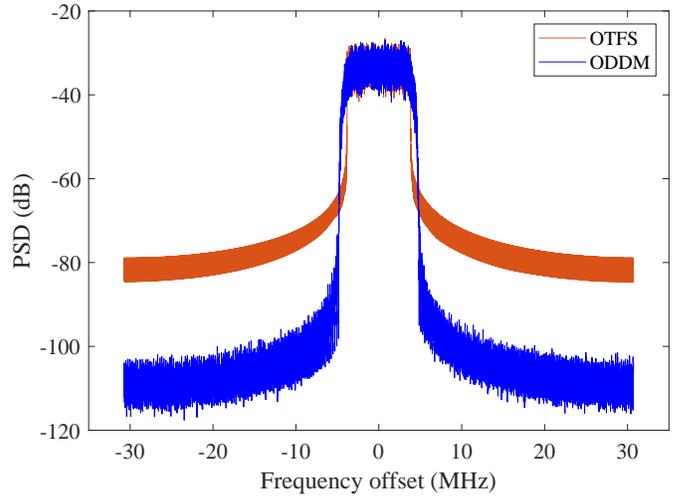}
	\caption{PSD comparison, $M=512$, $N=64$, $4$-QAM.}
	\label{powspec_comp}
	\vspace{-5mm}
\end{figure}

\begin{figure}
	\centering
	\includegraphics[width=8.9cm]{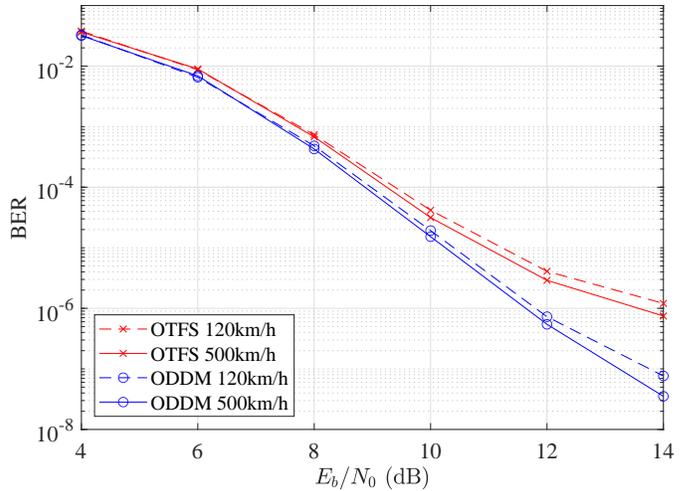}
	\caption{BER comparison, $M=512$, $N=64$, $4$-QAM.}
	\label{ber_comp_512_64}
	\vspace{-5mm}
\end{figure}

The comparison of power spectral density (PSD) is shown in Fig. \ref{powspec_comp}. There is no surprise to find that because of the $u(t)$-based pulse shaping, the ODDM has much lower OOB radiation than the OTFS, and up to $25$ dB improvement can be observed. As we mentioned before, the practical OFDM system cannot fully utilize all subchannels, leading to a reduced spectral efficiency considering  some null-subcarriers placed at the band edge.
This fact also applies to the OTFS, which however may become an issue for practical systems. Because the information-bearing signals are modulated in the DD domain, it is still unclear that how to arrange them in the DD domain to achieve unloaded edge subcarriers in the TF domain after the ISFFT.
A possible solution is to reduce $M$, which however is the number of subcarriers that is usually a power of $2$ in practice to exploit low-complexity fast Fourier transform and therefore cannot be changed freely. On the other hand, the proposed ODDM scheme can fully utilize all subchannels. By tuning the roll-off factor, a trade-off between the bandwidth and OOB radiation can be struck to achieve the desirable spectral efficiency. Furthermore, since $M$ now is the number of ODDM symbols rather than the number of subcarriers, it can be chosen flexibly to adjust the bandwidth together with the roll-off factor.

We also evaluate the BER performance of the uncoded ODDM and OTFS modulations, both with MP detection. For the ODDM, the MP detection is based on the DD domain channel matrix $\mathbf H$ in (\ref{H}). Fig. \ref{ber_comp_512_64} shows the BER performance with $4$-QAM signal, $M=512$, $N=64$. At low to medium SNRs, the ODDM achieves similar performance to the OTFS in various high mobility channels with UE speeds of $120$ km/h and $500$ km/h, which means they both have robust performance against the channel Doppler effect. At high SNRs, ODDM outperforms the OTFS by around $2$ dB at the BER of $1\times 10^{-6}$, thanks to the matched filtering and the exploiting of the exact DD domain channel input-output relation in the detection.

\section{Conclusion}
We studied the problem of MC modulation on DD plane, and reveal the link between the DD plane MC modulation and the conventional SMT modulation.
We then proposed the ODDM modulation, whose staggered upsampled-OFDM representation in digital domain was presented.
The transmit pulse of the ODDM was also proposed, and its orthogonality with respect to the DD  plane's resolutions was proved. By virtue of this favorable orthogonality, we derived an exact DD domain channel input-output relation of the proposed ODDM, where the equivalent DD domain channel matrix has an elegant block-circular-like structure. Finally, because of the perfect coupling between the modulated signal and the DD channel, the superior performance of the ODDM over the OTFS in terms of OOB radiation and BER was demonstrated by simulations.


%

\appendices

\section{Proof of the orthogonality of $u(t)$}
Since the real-valued filter $a(t)$ only has support on $\{-\frac{QT}{M}, \frac{QT}{M}\}$ and $Q \ll \frac{M}{2}$, for $|m| \le M-2Q$,
the ambiguity function of $u(t)$ is given by
\begin{align}
	 & A_{u,u}\left(m\frac{T}{M},n\frac{1}{NT}\right) = \int u(t)u\left(t-m\frac{T}{M}\right)e^{-j\frac{2\pi n}{NT}(t-m\frac{T}{M})} dt, \nonumber                                                                                     \\
	 & = \sum_{\dot n=0}^{N-1} \int_{\dot nT-\frac{QT}{M}}^{\dot nT+\frac{QT}{M}} a(t-\dot nT)a\left(t-\dot nT-m\frac{T}{M}\right) e^{-j\frac{2\pi n}{NT}(t-m\frac{T}{M})} dt, \nonumber                                               \\
	 & = e^{j\frac{2\pi m n }{MN}}\sum_{\dot n=0}^{N-1} e^{-j\frac{2\pi n \dot n }{N}} \int_{-\frac{QT}{M}}^{\frac{QT}{M}} a(\dot \tau)a\left(\dot \tau-m\frac{T}{M}\right)  e^{-j\frac{2\pi n \dot \tau}{NT}} d\dot \tau. \label{au2}
\end{align}
Because $\sum_{\dot n=0}^{N-1} e^{-j\frac{2\pi n \dot n }{N}} =0$ for $n\ne 0$, (\ref{au2}) becomes 
\vspace{-2mm}
\begin{align}\label{au3}
	A_{u,u}\left(m\frac{T}{M},n\frac{1}{NT}\right) & = N \delta(n) \int_{-\frac{QT}{M}}^{\frac{QT}{M}} a(\dot \tau)a\left(\dot \tau-m\frac{T}{M}\right) d\dot \tau \nonumber, \\
	                                               & = \delta(m) \delta(n).
\end{align}
\setlength{\belowdisplayskip}{0pt}
When $M-2Q<m\le M-1$, we can let $\dot m= m-M$. Then, notice that $0< |\dot m|<2Q\le M-2Q$, we have
\begin{align}
	 & A_{u,u}\left(m\frac{T}{M},n\frac{1}{NT}\right) = \int u(t)u\left(t-m\frac{T}{M}\right)e^{-j\frac{2\pi n}{NT}(t-m\frac{T}{M})} dt, \nonumber                                                                                               \\
	 & \stackrel{(a)}{=} \sum_{\dot n=1}^{N-1} \int_{\dot nT-\frac{QT}{M}}^{\dot nT+\frac{QT}{M}} a(t-\dot nT)a\left(t+T-\dot nT-m\frac{T}{M}\right) \nonumber                                                                                   \\
	 & \quad \quad \times e^{-j\frac{2\pi n}{NT}(t+T-m\frac{T}{M})} dt, \nonumber                                                                                                                                                                \\
	 & = \sum_{\dot n=1}^{N-1} \int_{\dot nT-\frac{QT}{M}}^{\dot nT+\frac{QT}{M}} a(t-\dot nT)a\left(t-\dot nT-\dot m\frac{T}{M}\right)  e^{-j\frac{2\pi n}{NT}(t-\dot m\frac{T}{M})} dt, \nonumber                                              \\
	 & = e^{j\frac{2\pi \dot m n }{MN}}\sum_{\dot n=1}^{N-1} e^{-j\frac{2\pi n \dot n }{N}} \int_{-\frac{QT}{M}}^{\frac{QT}{M}} a(\dot \tau)a\left(\dot \tau-\dot m\frac{T}{M}\right)  e^{-j\frac{2\pi n \dot \tau}{NT}} d\dot \tau, \label{au4}
\end{align}
where $\stackrel{(a)}{=}$ is due to the fact that the pulses $a(t-\dot nT)$ and $a(t-\dot nT-m \frac{T}{M})$ do not overlap but $a\left (t-\dot nT\right)$ and $a\left(t+T-\dot nT-m \frac{T}{M}\right) $ overlap for these $m$ values. When $n=0$, (\ref{au4}) becomes
\setlength{\belowdisplayskip}{5pt}
\vspace{-1mm}
\begin{align*}
	A_{u,u}\left(m\frac{T}{M},n\frac{1}{NT}\right) = (N-1) \int_{-\frac{QT}{M}}^{\frac{QT}{M}} a(\dot \tau)a\left(\dot \tau-\dot m\frac{T}{M}\right)  d\dot \tau = 0.
\end{align*}
When $n\ne 0$, for $\dot \tau \in \left(-\frac{QT}{M},\frac{QT}{M}\right)$, because of
$\frac{2\pi n \dot \tau}{NT} \in ( \frac{-2\pi n Q}{MN},\frac{2\pi n Q}{MN} )$ and
$2Q \ll M$, we have $e^{\frac{j2\pi n \dot \tau}{NT}} \approx 1$. Bearing in mind that $\sum_{\dot n=1}^{N-1} e^{-j\frac{2\pi n \dot n }{N}}=-1$ for $n\ne 0$, (\ref{au4}) then becomes
\vspace{-1mm}
\begin{align*}
	A_{u,u}\left(m\frac{T}{M},n\frac{1}{NT}\right) \approx -e^{j\frac{2\pi \dot m n }{MN}} \int_{-\frac{QT}{M}}^{\frac{QT}{M}} a(\dot \tau)a\left(\dot \tau-\dot m\frac{T}{M}\right)  d\dot \tau =  0,
\end{align*}
where the approximation error is negligible by further taking the shape of $a(t)$ into account. Meanwhile, for $-M+1 \le m <-M+2Q$, we can have results similar to (\ref{au4}), with different $\dot m$ and the corresponding different range for the summation indexed by $\dot n$. The combination of (\ref{au3}) and (\ref{au4}) completes the proof.

\ifCLASSOPTIONcaptionsoff
	\newpage
\fi




\vspace{-3mm}
\bibliographystyle{IEEEtran}
\bibliography{otfs}

\end{document}